\documentstyle[psfig]{basi}
\begin{document}
\title[Tiny supernovae in radio wavelengths]{ Baby supernovae through the 
looking glass at long wavelengths
}
\author[Poonam Chandra \& Alak Ray]%
       {Poonam Chandra$^{1,2}$, Alak Ray$^2$ \\ 
$^1$ Joint Astronomy Programme, IISc, Bangalore 560012\\
$^2$ Tata Institute of Fundamental Research, Mumbai 400005\\
}
\maketitle
\label{firstpage}
\begin{abstract}
We emphasize the importance of observations of young supernovae
in wide radio band. 
We argue on the basis of observational results 
that only high or only low frequency data is not 
sufficient to get full physical picture of the shocked plasma.
In the SN 1993J composite spectrum obtained with Very Large Array (VLA)
 and Giant Metrewave Radio Telescope (GMRT),
around day 3200, we see observational evidence of
synchrotron cooling, which leads us to the direct determination of the 
magnetic field independent of the equipartition assumption, as
well as the relative strengths of the magnetic field and relativistic
particles energy densities. 
The GMRT low frequency SN 1993J light curves suggests the modification
in the radio emission models developed on the basis of VLA data
alone. The composite radio spectrum of SN 2003bg on day
350 obtained with GMRT plus VLA strongly supports
the internal synchrotron self absorption as the dominant absorption mechanism.
 
\end{abstract}

\begin{keywords}
Core collapse supernovae -- SN1993J, SN2003bg, synchrotron losses,
acceleration mechanism, radio emission 
\end{keywords}

\section{Introduction}

Radio emission from supernovae is argued to
be due to the synchrotron emission from the forward shocked shell 
due to the relativistic electrons in presence of magnetic field.
The radio emission is initially absorbed from the 
external medium (free-free absorption i.e. FFA) or through 
synchrotron self absorption (SSA), i.e. an internal process. 

The most critical parameter which affects the synchrotron emission is
the magnetic field. The  
magnetic fields in a few supernovae have been 
estimated indirectly by assuming equipartition 
between relativistic energy density and the magnetic energy density. 
The magnetic field in the shocked plasma 
in supernovae is enhanced due to hydrodynamic instabilities 
in the plasma. While it is plausible, there is no
convincing argument for the equipartition assumption. 
In many classical radio sources, such as supernova remnants (SNRs) like
the Crab or Cassiopeia A, or in luminous radio galaxies, the radio spectral
index is found to steepen at high frequencies
(see e.g. Kardashev 1962). 
This is due to the
so called synchrotron aging of the source, as during the lifetime of
the source, electrons with high enough energies in a homogeneous magnetic
field will be depleted due to efficient synchrotron radiation compared
with the ones with lower energies.
An observation of a synchrotron break can yield a measurement
of the magnetic field  {\it independent
of the equipartition argument} if the {\it age} of the source {\it is known}.

We emphasize here  the need for broad-band  radio observations
of SNe to address the above issues. 
Combining radio data from a high frequency 
VLA and low frequency 
GMRT can offer such opportunities.
We discuss two supernovae in this context
- an eleven years old type IIb supernova SN 1993J and a one year 
old type Ic supernova SN 2003bg. 

\section{SN 1993J in M81}
SN 1993J exploded on March 28, 1993 in M81 (3.6 Mpc). 
The early spectrum of SN1993J showed the
characteristic hydrogen line signature of type II SNe,
but subsequently made a
transition to hydrogen-free, He-dominated type Ib SNe,
hence classified as type IIb SN.
\subsection{Observations}
We observed SN 1993J with the 
GMRT
in 610 and 235 MHz wavebands on 2001 Dec 30
and in 1420 MHz band on 2001 Oct 15.
 We combined this
dataset with the high frequency VLA observations 
provided by C. Stockdale, K. Weiler et al. 
in 22.5, 14.9, 8.4, 4.8
and 1.4 GHz wavebands observed on 2002 Jan 13 (See Table \ref{tab:1}).
The data was analyzed using Astronomical Image Processing System (AIPS).
More details of observations, data analysis 
are described in Chandra et al. 2004a.

\begin{table}
\caption{Observations of the spectrum of SN 1993J on day 3200
\label{tab:1}}
\begin{tabular}{cccccc}
\hline\hline
Date of  & Telescope & Days since & Frequency &  Flux density & rms\\
Observation& &  explosion & in GHz  & mJy   & mJy \\
\hline\hline
Dec 31,01 & GMRT & 3199 & 0.239  & 57.8 $\pm$7.6   & 2.5\\
Dec 30,01 & GMRT & 3198 & 0.619  & 47.8 $\pm$5.5   & 1.9\\
Oct 15,01 & GMRT & 3123 & 1.396  & 33.9 $\pm$3.5   & 0.3\\
Jan 13,02 & VLA & 3212 & 1.465  & 31.44$\pm$4.28  & 2.9\\
Jan 13,02 & VLA & 3212 & 4.885  & 15   $\pm$0.77  & 0.19\\
Jan 13,02 & VLA & 3212 & 8.44   & 7.88 $\pm$0.46  & 0.24\\
Jan 13,02 & VLA & 3212 & 14.965 & 4.49 $\pm$0.48  & 0.34\\
Jan 13,02 & VLA & 3212 & 22.485 & 2.50 $\pm$0.28  & 0.13\\
\hline\hline
\end{tabular}
\end{table}

\subsection{Modeling the composite radio spectrum}
	
The GMRT and VLA combined spectrum on day 3200 suggests the break in the 
spectral index in the optically thin part of the spectrum (Fig. 1).
The break in the
spectrum occurs at $4.02\pm 0.19$ GHz with the steepening of spectral 
index by 0.6. 
This variation in spectral index is 
consistent with that
predicted from the synchrotron cooling effect
with continuous injection (Kardashev 1962). If we were to
model the low frequency data with a turnover due to SSA, then
{\it under the assumption of equipartition} we would
obtain a field of
$B_{eq}=38.3\pm 17.1$ mG (see, however next section).

\begin{figure}
\centerline{
\psfig{figure=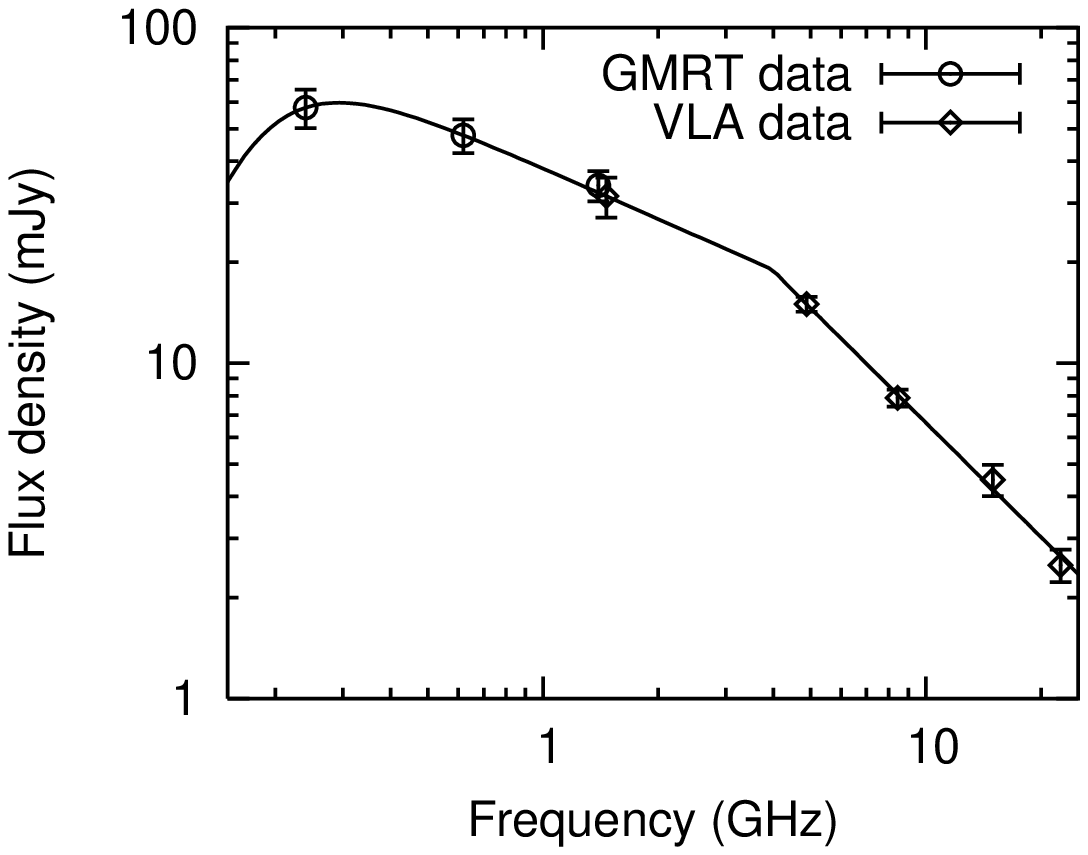,width=6.8cm}
\psfig{figure=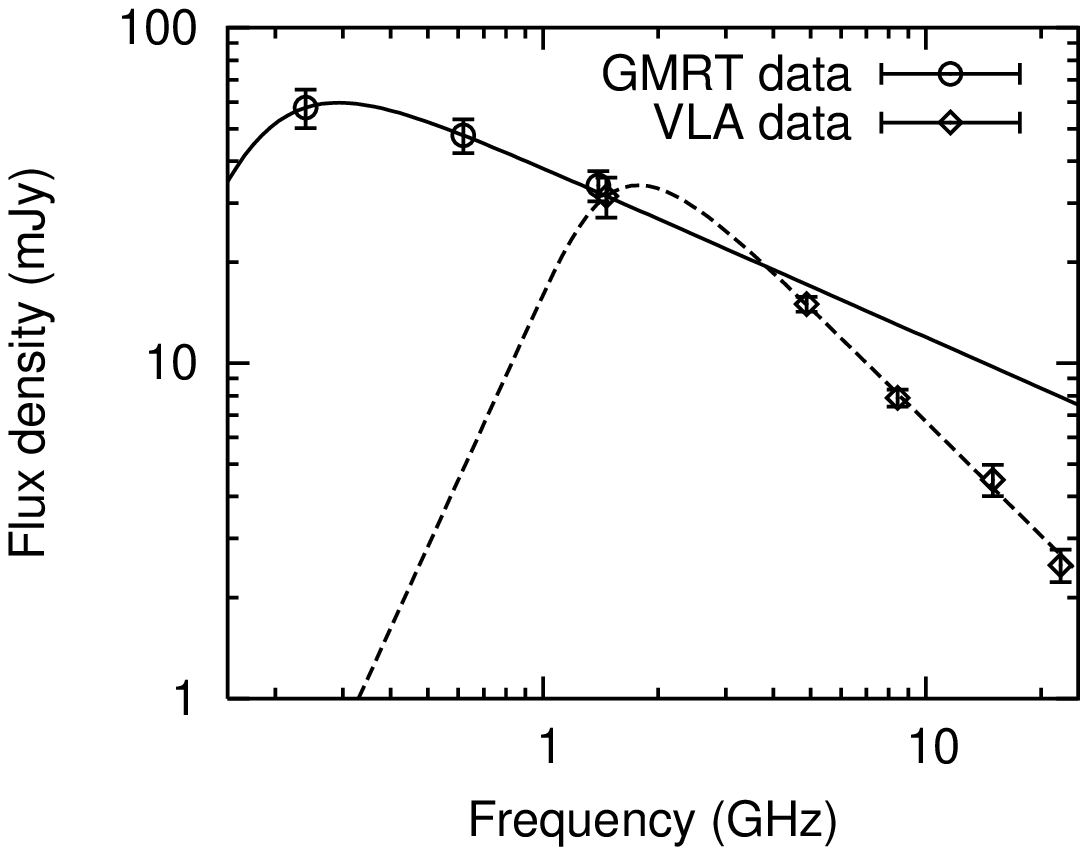,width=6.8cm}}
{\small Figure 1. {\it Left panel}: Combined GMRT plus VLA spectrum of
SN 1993J on day 3200 with SSA model (solid line)
with a break in the spectral index at 4 GHz.
At low frequencies, the spectral emission index $\alpha$ is 0.51
before break and after the
break, in high frequency regime, it is 1.13.
{\it Right panel}: 
Wide band spectrum of SN 1993J. Synchrotron self absorption
fit to "only" low frequency data
(solid line) and "only" high frequency data (dashed line).
}
\end{figure}

\subsection{Synchrotron aging and determination of the magnetic field}

%
The lifetime of the relativistic electrons undergoing synchrotron losses is
given as
\begin{equation}
\tau = E/[-{(dE/dt)}_{Sync}]=1.43 \times 10^{12} B^{-3/2} {\nu}^{-1/2}\; {\rm sec}
\end{equation}
Here we use  ${B_{\perp}}^2={(B\, {\rm Sin \theta})}^2 =(2/3)B^2$.
The above expression
is implicitly a function of time, since the magnetic field in the
region of emission itself
changes with time as the supernova shock moves out farther
into the circumstellar plasma.
The time variation of the synchrotron break frequency can be obtained
by setting: $\tau = t$, whence,

\begin{equation} \label{sync}
{{\nu}_{break}} = {(t/1.43 \times 10^{12})}^{-2} B^{-3}
=2 \times 10^{24} B_0^{-3}\, t
\; {\rm Hz}
\end{equation}

Here we use $B=B_0/t$ (Fransson \& Bjornsson 1998).
From the above eqn.
(and using ${\nu}_{break}=5.12\times 10^{18} B
 E_{break}^2 \;{\rm Hz}$ (Pacholczyk 1969))
and with break frequency 4 GHz, we get magnetic
field $B=0.19$ G for $t=3200$ days.
However this estimate of the $B$
does not account for
other processes like diffusive shock acceleration (Fermi mechanism)
and adiabatic losses, likely to be important for a young
supernova. 

We derive below the magnetic field under cumulative
effect of all these processes. The adiabatic losses
will be given by
${dE/dt}_{Adia}=-(V/R)E=-E/t$. Here
$V$ is the expansion velocity, i.e. the
ejecta velocity and $R$ is the radius of
the spherical shell.

In supernovae, diffusive mechanism is assumed to be the main acceleration
mechanism (Fransson \& Bjornsson 1998, Ball \& Kirk 1992). In this process
electrons gain energy every time they cross the shock front either from
upstream to downstream or vice versa.
The average fractional momentum gain per shock crossing or recrossing is:
$\Delta  =(4(v_1-v_2)/3v) $ and the average time taken to
perform one such cycle is (Ball \& Kirk 1992, Drury 1983),
$t_c=4{\kappa}_{\perp}(1/v_1+1/v_2)/v$.
Here $v$ is the test particle velocity, $v_1$ is the upstream velocity
and $v_2$ is the downstream velocity, and
${\kappa}_{\perp}$ is the spatial diffusion coefficient
of the test particles across the ambient magnetic field, when
the shock front is quasi-perpendicular to the field.
In the rest frame of shock front,
$v_1=V$ and $v_2=v_1/4=V/4$ (for compression factor of 4).
The break in the spectrum will occur for those
electron energies for which the time scales for the
cumulative rate of change of electron energy
due to synchrotron cooling plus adiabatic losses
and gain through diffusive acceleration
becomes comparable to the life time of the supernova (Kardashev 1962).
Lifetime of electrons  for the cumulative energy loss rate is
\begin{equation}
\tau = \frac{E}{{(dE/dt)}_{Total}}=\frac{E}{ (R^2 t^{-2}/20 {\kappa}_{\perp})E
-b B^2  E^2- t^{-1}E}
\end{equation}
where the first term in the denominator is the acceleration term, 
the second term is the familiar synchrotron loss term with
$b=1.58 \times 10^{-3}$ and the third term is due to
adiabatic losses. Setting
the life time $\tau=t$, break frequency can be derived as:
\begin{equation}
\label{synchr_freq}
{\nu}_{break}=\frac{2\times 10^{24}}{{B_0}^3}
{\left[\frac{R^2}{20 {\kappa}_{\perp}}t^{-1/2}-2\,t^{1/2}\right]}^2
 \,{\rm Hz}
\end{equation}

The value of ${\kappa}_{\perp}$ is used as 
$2.96 \times 10^{24}$ cm$^2$s$^{-1}$ (see Chandra et al. 2004b). 
Using size of the supernova $R=2.65\times 10^{17}$ cm from VLBI
observations (Bartel et al. 2002), we obtain magnetic field
$B=0.33\pm0.01 $ Gauss, from the observationally determined break.
On the other hand, from the best fit in SSA, the magnetic
field under equipartition assumption is $B_{eq}=38\pm17$ mG.
Comparison of the two magnetic field determines the 
value of the equipartition fraction between relativistic energy of
particles and magnetic field energy. Equipartition fraction
$a=U_{rel}/U_{mag}$  varies with magnetic field $B$ as
$a = (B/B_{eq})^{-(2 \gamma +13)/4}$ (Chevalier 1998).
Therefore, the fraction $a$ ranges between $ 8.5 \times
10^{-6}\,-\, 4.0 \times 10^{-4}$ with a central value (corresponding
to $B_{eq}=38$ mG) of $a=1.0 \times 10^{-4}$ on day 3200.
This very low value of the equipartition fraction 
suggests that the plasma is
heavily dominated by the magnetic energy density, and electron 
acceleration to relativistic energies is inefficient.

\subsection{Role of acceleration}

It is argued here that acceleration may play a role in determination
of magnetic field from the synchrotron 
cooling break when the acceleration region and
the synchrotron loss regions are overlapping. 
However, the extent to which these two regions overlap near the interaction 
shell strongly distorted by fingers of hydrodynamic instability is not 
a priori clear.
The experimental trend of shift in the break frequency with time may
determine this issue. If acceleration processes are important then 
break frequency will tend to shift to lower frequency with time 
($\nu_{break} \propto t^{-1}$),
whereas break frequency will shift to higher frequency with time
($\nu_{break} \propto t$) 
if they are not important (Eq. \ref{synchr_freq}). 
In any case, our conclusion that the plasma is dominated by magnetic energy
density remains unaffected.

\section{ SN 1993J light curves at low frequencies}

Weiler et al. 2002
 provided the detailed modeling of SN 1993J at all epochs
based on high frequency VLA data. 
We extrapolate this model to 1420, 610, 325 and
243 MHz frequencies at GMRT observation epochs with the above
parameters. Fig 2. shows this extrapolated light
curves for the SN at 1420, 610, 325 and 243 MHz and our corresponding GMRT
data points at the respective frequencies. It is evident that the
free-free model described above overpredicts the flux densities at low
frequencies. In fact lower the frequencies, more significant is the
departure from the standard free-free model. This indicates that
the optical depths fitted using high frequency datasets,
simply extrapolated to low frequencies with the
dependence $\tau \propto {\nu}^{-2.1}$ are not sufficient to account for the
required absorption.
 One needs to incorporate some additional frequency dependent
opacity at low frequencies, which can compensate for the difference
between the model light curves and actual data.

\begin{figure}
\hbox{
\psfig{figure=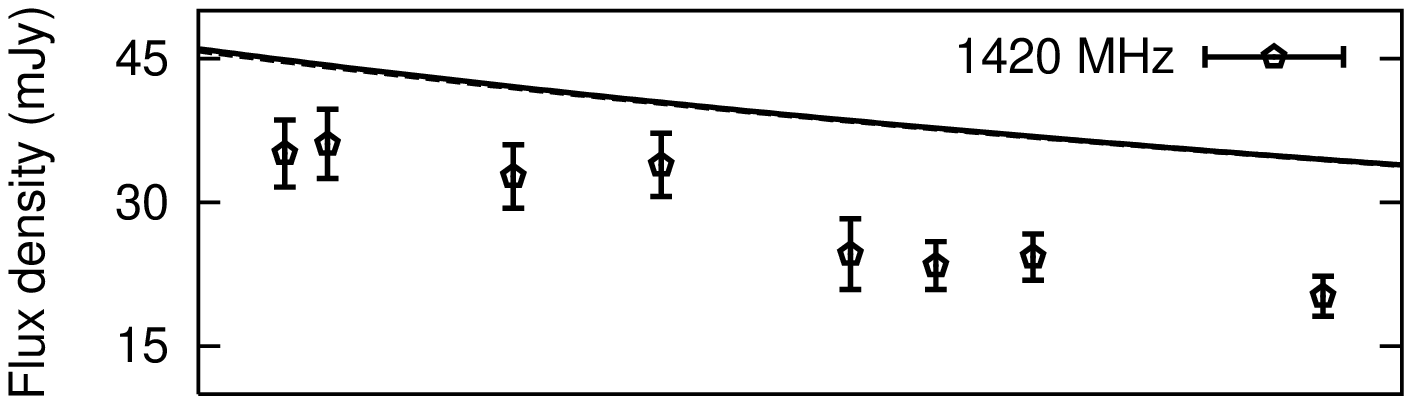,width=2.6in}
\hspace{0.2cm}
\psfig{figure=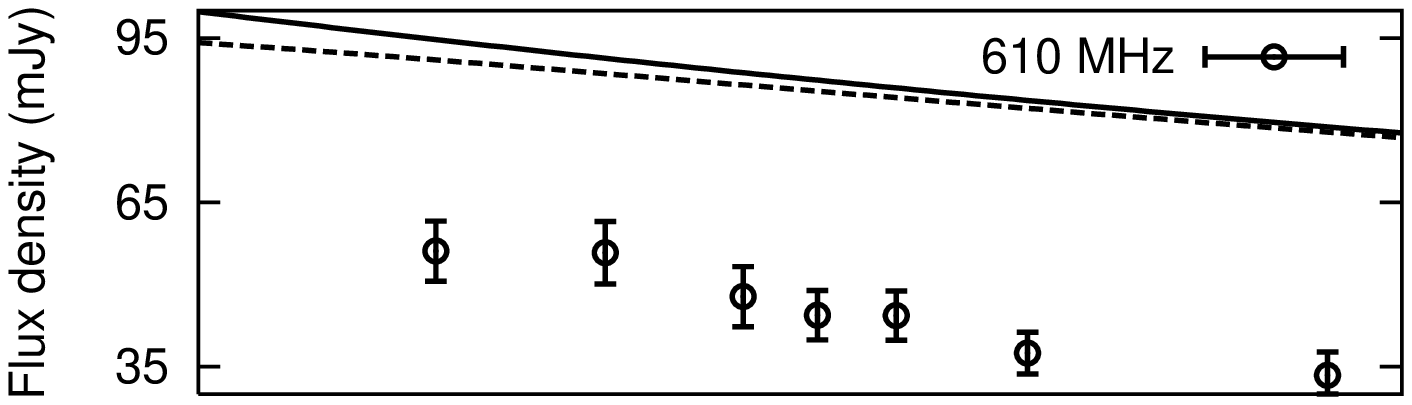,width=2.6in}} 
\hbox{
\psfig{figure=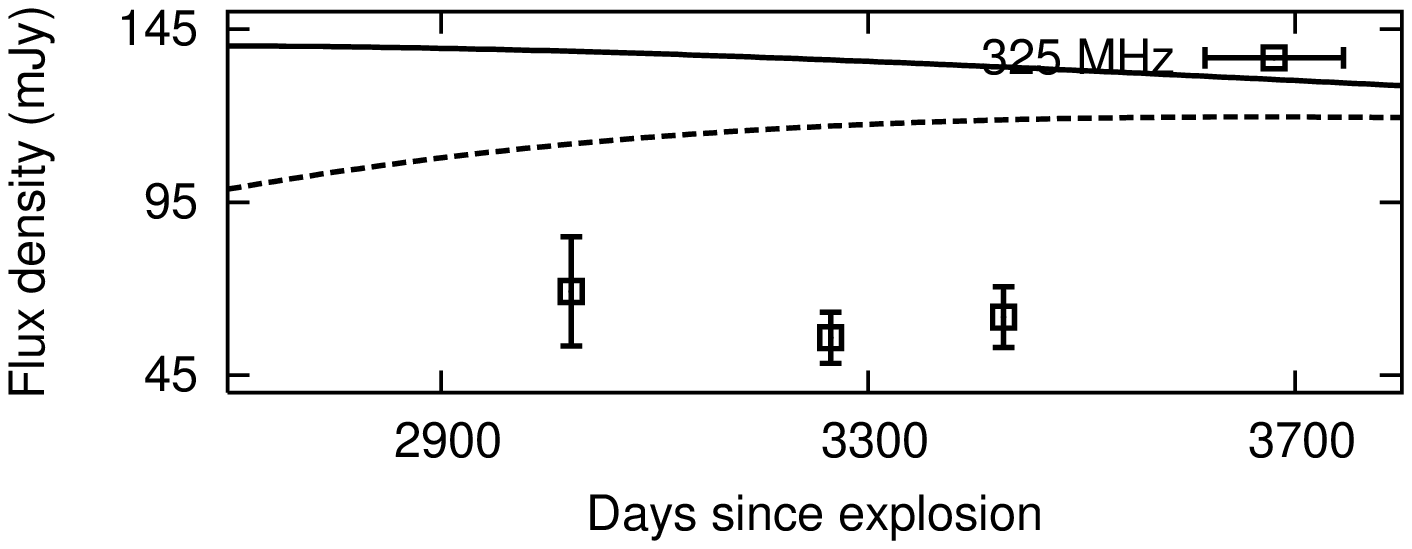,width=2.6in}
\hspace{0.2cm}
\psfig{figure=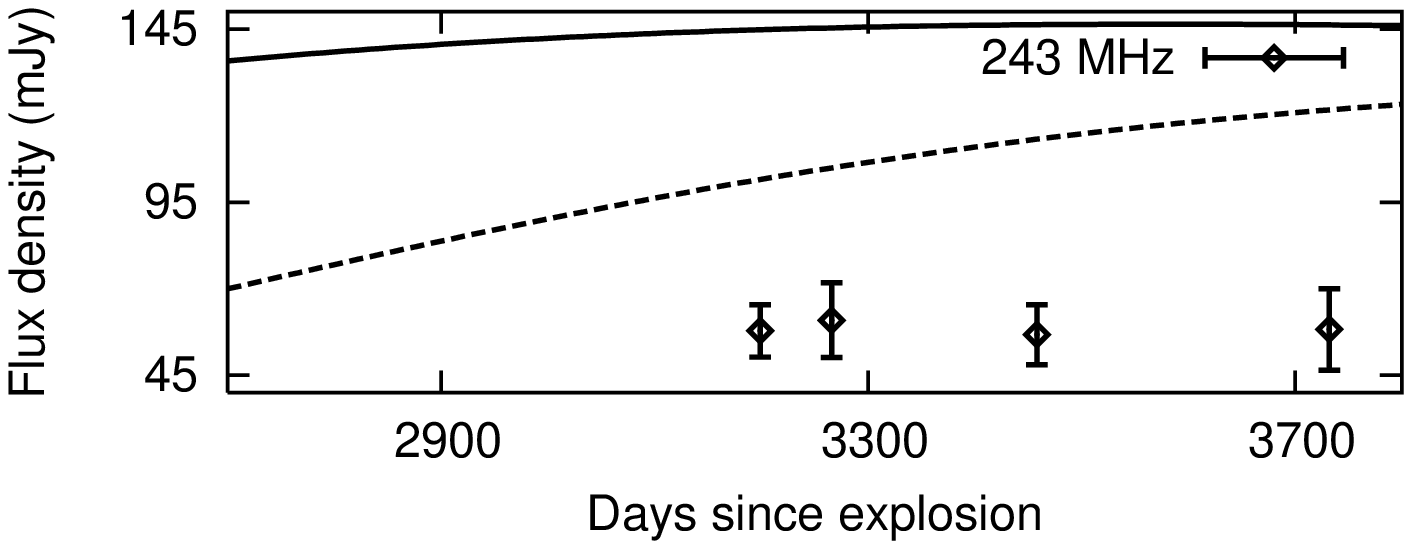,width=2.6in}}
{\small
Figure 2.: 
Comparison of low frequency
data to the predictions of the models obtained by fitting
high frequency fluxes of SN 1993J. The  solid lines in
all the three plots are Weiler et al's (2002)  model extrapolated
to lower frequencies. Dashed lines are the flux density plot after
incorporating the SSA optical depth in the Weiler's free-free model.
}
\end{figure}

\section{SN 2003bg in MCG -05-10-015}

SN 2003bg is type Ic supernova in MCG -05-10-015 (19 Mpc). It was discovered
on 2003 Feb 25, most likely two weeks after the explosion (based
on the spectral chronometers).
It was devoid of Hydrogen and Helium in the optical spectra, hence 
classified as type Ic supernova.

\subsection{Observations}

We observed SN 2003bg with GMRT in 1280 MHz band on 2003 Feb 2
and then in 610 MHz band on 2003 Feb 5 and in 325 MHz band on 2003 Feb 8. 
On our request A. Soderberg and S. Kulkarni  
observed the SN at high VLA frequencies on 2003 Feb 8, thus obtaining
the wide band radio spectrum all the way from 0.3 GHz up to 44 GHz
(Tab. 2). Fig. 3
shows the 1280 MHz band GMRT radio map of SN 2003bg.

\subsection{Modeling the composite radio spectrum}

Radio emission in SNe is usually absorbed in the early stages. 
It can be
either due to FFA or SSA. The data in the optically thick part of the
spectrum can distinguish between the two 
because of their varied dependence on frequency.
We fit both homogeneous 
FFA and SSA models (Chevalier 1998) to the spectrum (see Fig. 3).
The 610 MHz data point
 clearly discriminates the FFA model and favors SSA model.
From the SSA fit to the data under equipartition assumption,
we find the values of the following parameters: emission spectral
index ($\alpha = 3$), size of SN 2003bg ($R= (9.98 \pm 0.43) \times 
10^{16}$ cm), expansion speed ($ v = 33,000\,{\rm km \, s^{-1}}$),
and magnetic field ($ B = 0.18 \pm 0.03$ G). 

We notice that 44 GHz data point is not falling on the powerlaw in the
optically thin part and it suggests 
a break somewhere between 22 GHz to 44 GHz. 
The break may be due to the synchrotron aging. Since we have only one 
data point, we cannot draw any conclusive evidence and 
it will be useful to have the spectrum extended beyond 44 GHz.
If there is indeed a break due to synchrotron cooling, it 
will determine the magnetic field directly, independent of 
equipartition assumption, as was in the case of SN 1993J.

\begin{table}
\caption{Observations of the spectrum of SN 2003bg on day 350
\label{tab:2}}
\begin{tabular}{cccccc}
\hline\hline
Date of  & Telescope & Days since & Frequency &  Flux density & rms\\
Observation& &  explosion & in GHz  & mJy   & mJy \\
\hline\hline
Feb 08,04 & GMRT & 363 & 0.325  & 57.8 $\pm$7.6   & 2.5\\
Feb 05,04 & GMRT & 360 & 0.619  & 47.8 $\pm$5.5   & 1.9\\
Feb 02,04 & GMRT & 357 & 1.280  & 33.9 $\pm$3.5   & 0.3\\
Feb 08,04 & VLA & 363 & 1.465  & 23.88$\pm$0.4  & $-$ \\
Feb 08,04 & VLA & 363 & 4.885  & 30.88 $\pm$0.08  & $-$\\
Feb 08,04 & VLA & 363 & 8.44   & 18.96 $\pm$0.07  & $-$\\
Feb 08,04 & VLA & 363 & 14.97 & 9.65 $\pm$0.17  & $-$ \\
Feb 08,04 & VLA & 363 & 22.5 & 6.56 $\pm$0.08  & $-$\\
Feb 08,04 & VLA & 363 & 44.3 & 2.23 $\pm$0.26  & $-$ \\
\hline\hline
\end{tabular}
\end{table}

\begin{figure}
\centerline{
\psfig{figure=f3a.eps,width=5.0cm}
\psfig{figure=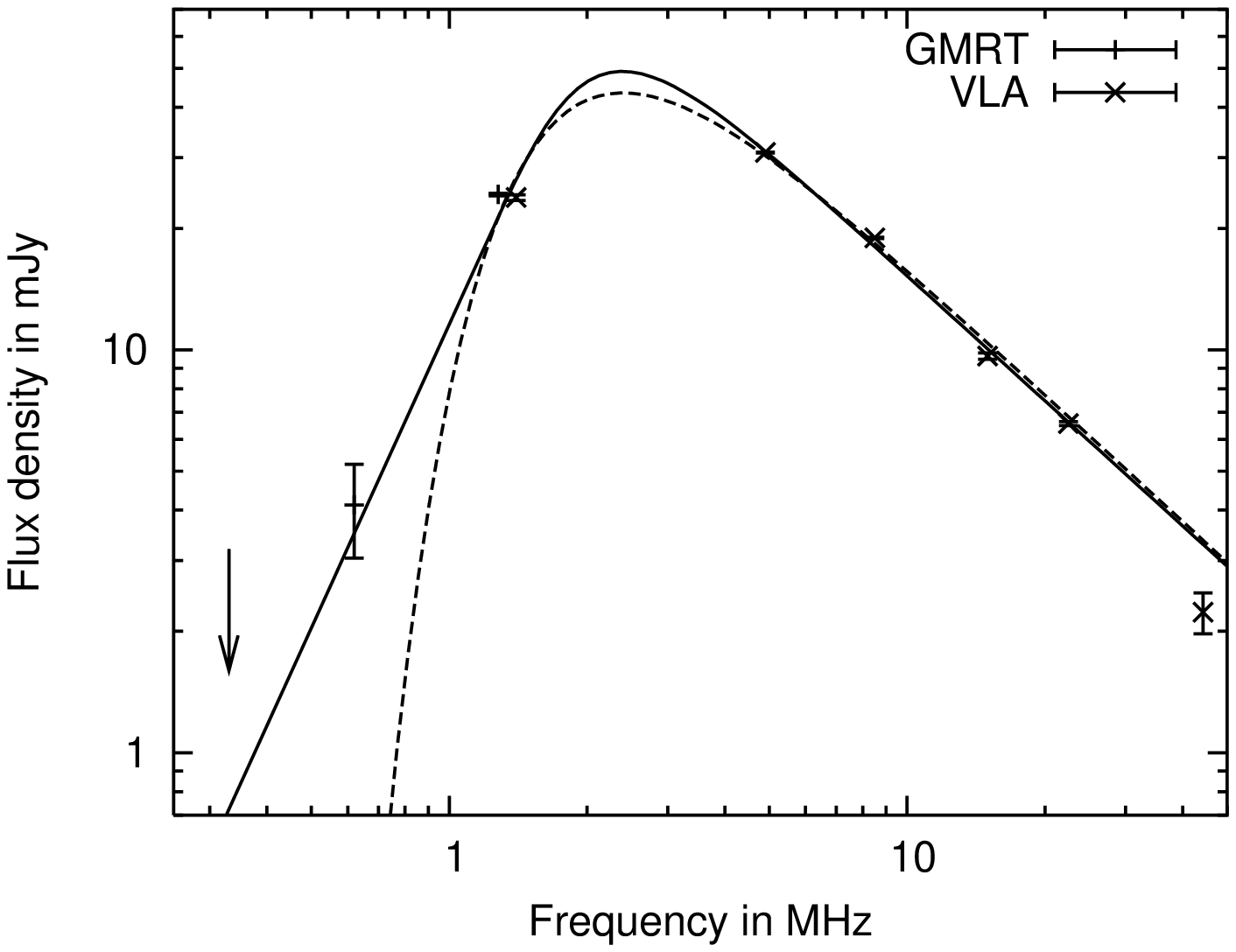,width=7.8cm}}
{\small Figure 3. {\it Left panel}: 
GMRT radio map of SN 2003bg at 1280 MHz frequency observed on Feb 02, 2004.
{\it Right panel}:
Synchrotron self absorption (solid line) and free-free absorption (dashed 
line) fits to the SN 2003bg combined GMRT plus VLA spectrum.
}
\end{figure}

\section{Discussion and Conclusions}

In right panel of Fig. 1 
we show a comparison of SSA 
model (with a single optically-thin power-law index) fitted
only to the low frequency data (0.22 GHz to 1.4 GHz) versus such a model
fit obtained with only the higher frequency data (1.4 GHz to 22.5 GHz).
This comparison shows that while the model fitted only to
the low frequencies over-predicts
the flux density at high frequencies, the model
fitted only to high frequencies on the other hand fails
to account for both synchrotron cooling break and
 seriously under-predicts the low frequency flux densities.
The comparison underscores the importance of broad band observations
for determining the physical processes taking place in the supernova.

The particle energy density is far below than the magnetic energy density
(by a factor of 1/10000) for SN 1993J, although there are indications that in 
SNe like SN 1998bw/GRB980425, there may exist this 
equipartition (Kulkarni et al. 1998).
Future studies of the equipartition factor for SN 1993J may indicate for the 
first time how particle acceleration efficiency in 
strongly turbulent magnetized plasma evolves with time
in the large magnetic-Reynolds and Reynolds numbers limit.
Light curves based on high frequency FFA models
 extrapolated to low frequencies overpredict the fluxes
at low frequencies. Some extra opacity is needed to incorporate the
difference. We added an extra opacity due to SSA
which also could not account for the required absorption (Fig. 2).
This suggests the low frequency opacity in SN 1993J is not the simple
extrapolation of high frequency opacity and it is likely that hitherto
unaccounted  absorption mechanisms are at work at low frequencies.

There could be a break in the radio spectrum of SN 2003bg.
We are planning simultaneous observations with GMRT along with VLA 
and ATCA. This will provide us the spectrum from 0.2 GHz to 80 GHz
and will establish whether the break is real or due to data artifact.
We also could discard the 
homogeneous free-free absorption model over the SSA model from the
low-frequency optically thin part of the spectrum.

%

We thank Dr. S. Bhatnagar for crucial help in the analysis of early GMRT
data for SN 1993J. We thank K. Weiler, C. Stockdale and their 
collaboration for kindly providing the VLA data for SN 1993J at high 
frequencies. We also thank A. Soderberg and S. Kulkarni
for observing SN 2003bg on our request and providing us the results
of observations.
We thank the staff of the GMRT, NCRA (TIFR)
 that made these observations possible. 
Poonam Chandra is a Sarojini Damodar International Fellow.

\end{document}